\documentclass[twocolumn]{aastex63}
\usepackage{chemformula}
\usepackage{float}

\shorttitle{Is the OPR of \ch{H2CO} Preserved?}
\shortauthors{J. Terwisscha van Scheltinga et al.}

\begin{document}

\title{The TW~Hya Rosetta Stone Project II: Spatially resolved emission of formaldehyde hints at low-temperature gas-phase formation}

\correspondingauthor{J. Terwisscha van Scheltinga}
\email{jeroentvs@strw.leidenuniv.nl}

\author[0000-0002-3800-9639]{Jeroen Terwisscha van Scheltinga}
\affiliation{Laboratory for Astrophysics, Leiden Observatory, Leiden University, PO Box 9513, 2300 RA Leiden, The Netherlands}
\affiliation{Leiden Observatory, Leiden University, PO Box 9513, 2300 RA Leiden, The Netherlands}

\author[0000-0001-5217-537X]{Michiel R. Hogerheijde}
\affiliation{Leiden Observatory, Leiden University, PO Box 9513, 2300 RA Leiden, The Netherlands}
\affiliation{Anton Pannekoek Institute for Astronomy, University of Amsterdam, Science Park 904, 1098 XH, Amsterdam, The Netherlands}

\author[0000-0003-2076-8001]{L. Ilsedore Cleeves}
\affiliation{Astronomy Department, University of Virginia, Charlottesville, VA 22904, USA}

\author[0000-0002-8932-1219]{Ryan A. Loomis}
\affiliation{National Radio Astronomy Observatory, 520 Edgemont Rd, Charlottesville, VA 22903, USA}

\author[0000-0001-6078-786X]{Catherine Walsh}
\affiliation{School of Physics and Astronomy, University of Leeds, Leeds LS2 9JT, UK}

\author[0000-0001-8798-1347]{Karin I. {\"O}berg}
\affiliation{Center for Astrophysics $\vert$ Harvard \& Smithsonian, 60 Garden Street, Cambridge, MA 02138, USA}

\author[0000-0003-4179-6394]{Edwin A. Bergin}
\affiliation{Department of Astronomy, University of Michigan, 1085 South University Avenue, Ann Arbor, MI 48109, USA}

\author[0000-0002-8716-0482]{Jennifer B. Bergner}
\affiliation{University of Chicago, Department of the Geophysical Sciences, Chicago, IL 60637, USA}

\author[0000-0003-0787-1610]{Geoffrey A. Blake}
\affiliation{Division of Chemistry \& Chemical Engineering, California Institute of Technology, Pasadena CA 91125, USA}
\affiliation{Division of Geological \& Planetary Sciences, California Institute of Technology, Pasadena CA 91125, USA}

\author[0000-0002-0150-0125]{Jenny K. Calahan}
\affiliation{Department of Astronomy, University of Michigan, 1085 South University Avenue, Ann Arbor, MI 48109, USA}

\author[0000-0002-1917-7370]{Paolo Cazzoletti}
\affiliation{Leiden Observatory, Leiden University, PO Box 9513, 2300 RA Leiden, The Netherlands}

\author[0000-0001-7591-1907]{Ewine F. van Dishoeck}
\affiliation{Leiden Observatory, Leiden University, PO Box 9513, 2300 RA Leiden, The Netherlands}
\affiliation{Max-Planck-Institut f{\"u}r Extraterrestrische Physik, Giessenbachstra{\ss}e 1, D-85748 Garching bei M{\"u}nchen, Germany}

\author[0000-0003-4784-3040]{Viviana V. Guzm\'an}
\affiliation{Instituto de Astrofísica, Ponticia Universidad Católica de Chile, Av. Vicuña Mackenna 4860, 7820436 Macul, Santiago, Chile}

\author[0000-0001-6947-6072]{Jane Huang}
\altaffiliation{NHFP Sagan Fellow}
\affiliation{Center for Astrophysics $\vert$ Harvard \& Smithsonian, 60 Garden Street, Cambridge, MA 02138, USA}
\affiliation{Department of Astronomy, University of Michigan, 1085 South University Avenue, Ann Arbor, MI 48109, USA}

\author[0000-0003-0065-7267]{Mihkel Kama}
\affiliation{Department of Physics and Astronomy, University College London, Gower Street, London, WC1E 6BT, UK}
\affiliation{Tartu Observatory, University of Tartu, 61602 T\~{o}ravere, Estonia}

\author[0000-0001-8642-1786]{Chunhua Qi}
\affiliation{Center for Astrophysics $\vert$ Harvard \& Smithsonian, 60 Garden Street, Cambridge, MA 02138, USA}

\author[0000-0003-1534-5186]{Richard Teague}
\affiliation{Center for Astrophysics $\vert$ Harvard \& Smithsonian, 60 Garden Street, Cambridge, MA 02138, USA}

\author[0000-0003-1526-7587]{David J. Wilner}
\affiliation{Center for Astrophysics $\vert$ Harvard \& Smithsonian, 60 Garden Street, Cambridge, MA 02138, USA}




\begin{abstract}
Formaldehyde (\ch{H2CO}) is an important precursor to organics like methanol (\ch{CH3OH}). It is important to understand the conditions that produce \ch{H2CO} and prebiotic molecules during star and planet formation. \ch{H2CO} possesses both gas-phase and solid-state formation pathways, involving either UV-produced radical precursors or CO ice and cold ($\lesssim 20$ K) dust grains. To understand which pathway dominates, gaseous \ch{H2CO}'s ortho-to-para ratio (OPR) has been used as a probe, with a value of 3 indicating ``warm'' conditions and $<3$ linked to cold formation in the solid-state. We present spatially resolved ALMA observations of multiple ortho- and para-\ch{H2CO} transitions in the TW~Hya protoplanetary disk to test \ch{H2CO} formation theories during planet formation. We find disk-averaged rotational temperatures and column densities of $33\pm2$~K, ($1.1\pm0.1)\times10^{12}$~cm$^{-2}$ and $25\pm2$~K, $(4.4\pm0.3)\times10^{11}$~cm$^{-2}$ for ortho- and para-\ch{H2CO}, respectively, and an OPR of $2.49\pm0.23$. A radially resolved analysis shows that the observed \ch{H2CO} emits mostly at rotational temperatures of 30--40~K, corresponding to a layer with $z/R\ge0.25$. The OPR is consistent with 3 within 60~au, the extent of the pebble disk, and decreases beyond 60~au to $2.0\pm0.5$. The latter corresponds to a spin temperature of 12~K, well below the rotational temperature. The combination of relatively uniform emitting conditions, a radial gradient in the OPR, and recent laboratory experiments and theory on OPR ratios after sublimation, lead us to speculate that {\em gas-phase} formation is responsible for the observed \ch{H2CO} across the TW~Hya disk.
\end{abstract}



\section{Introduction} \label{sec:intro}

The incorporation of complex organic molecules (COMs) into forming planets is essential to solving the puzzle of life's origins \citep[e.g.,][]{2009_Herbst_vanDishoeck_ARA&A..47..427H}. The answer to how and where prebiotic molecules are formed is an important step in this investigation, and starts with the study of the chemical precursors of COMs. Even for one of the simplest COMs, methanol (\ch{CH3OH}), the origin of its precursor molecule formaldehyde (\ch{H2CO}) has yet to be fully constrained in protoplanetary disks \citep{2015_Loomis_DMTAU_ApJ...809L..25L, 2017_Oberg_H2CO_ApJ...839...43O}. Specifically, \ch{H2CO} presents a challenge in that it can potentially form via reactions in the gas phase and via formation in the ice mantles of cold grains, followed by non-thermal desorption or sublimation \citep[][]{2013_Qi_H2CO_N2H+_ApJ...765...34Q,2015_Loomis_DMTAU_ApJ...809L..25L,2017_Oberg_H2CO_ApJ...839...43O}. The relative occurrence of both paths is important, because they take place in different environments and thus contribute differently to the formation of methanol and other COMs. Furthermore, an unsolved question is whether the observed organic reservoir is close enough to the midplane where planets form. 

Solid-state formation of \ch{H2CO} starts with the hydrogenation of CO; further hydrogenation, though with a small barrier of 400--500 K, leads to efficient formation of \ch{CH3OH} \citep{1994_Hiraoka_hydro_CPL...229..408H, 2002_Hiraoka_hydro_ApJ...577..265H, 2002_Watanabe_hydro_ApJ...571L.173W, 2004_Hidaka_H2CO_ApJ...614.1124H, 2004_Watanabe_hydro_ApJ...616..638W, 2009_Fuchs_hydro_A&A...505..629F}. From \ch{CH3OH}, a complex and varied chemistry can be seeded by the subsequent formation of simple sugars and sugar alcohols like glycerol, an important building block for cell membranes \citep{2017_KoJu_COMs_MNRAS.467.2552C,2017_Gleb_Glycerol_ApJ...842...52F}. In contrast, gas-phase formation of \ch{H2CO} occurs most efficiently through the reaction between atomic oxygen and methyl radicals (\ch{CH3}) \citep{2002_Fockenberg_gasH2CO_JPCA..106.2924F,2006_Atkinson_gasH2CO_ACP.....6.3625A} as well as CH$_2$ and hydroxyl radicals (OH). Therefore, the gas-phase formation pathway is particularly efficient where these radicals can be generated, primarily in the UV irradiated surface where there is efficient photodesorption and photodissociation \citep{2002_Aikawa_warm_layer_A&A...386..622A, 2015_Loomis_DMTAU_ApJ...809L..25L}.

\begin{deluxetable*}{lcccccccc}
\tablecaption{ALMA Observations \label{tab:obs_data}}
\tablewidth{\textwidth}
\tabletypesize{\footnotesize}
\tablehead{
\colhead{ALMA}  & \colhead{Date}    & \colhead{Antennas}    & \colhead{Baselines}            & \colhead{On-source}   & \colhead{}        & \colhead{Calibrators} & \colhead{} \vspace{-2mm} \\
\colhead{Project code}      & \colhead{}        & \colhead{}        & \colhead{[m]}         & \colhead{[minutes]}   & \colhead{Bandpass}  & \colhead{Phase}       & \colhead{Flux}
}
\startdata
2013.1.00114.S\tablenotemark{\footnotesize a} & 2014 Jul 19 \tablenotemark{\scriptsize1} & 32 & 34-650 & 42.0 & J1037--2934 & J1037--2934 & Pallas \\
2016.1.00311.S\tablenotemark{\footnotesize b} & 2016 Dec 16 \tablenotemark{\scriptsize2} & 45 & 15-449 & 23.9 & J1037--2934 & J1037--2934 & J1037--2934 \\
 & 2017 Feb 01 \tablenotemark{\scriptsize3} & 41 & 14-256 & 28.3 & J1058+0133 & J1037--2934 & J1107--4449 \\
 & 2017 Apr 08 \tablenotemark{\scriptsize4} & 40 & 15-379  & 28.8 & J1037--2934 & J1037--2934 & J1058+0133 \\
 & 2017 May 05 \tablenotemark{\scriptsize2} & 45 & 16-1120 & 39.3 & J1037-2934 & J1037--2934 & J1107--4449 \\
 & 2017 May 07 \tablenotemark{\scriptsize2} & 51 & 16-1079 & 39.3 & J1037-2934 & J1037--2934 & J1107--4449 \\
 & 2017 May 21 \tablenotemark{\scriptsize4} & 45 & 15-1097 & 47.8 & J1037--2934 & J1037--2934 & J1037--2934 \\
 & 2018 Jan 23 \tablenotemark{\scriptsize3} & 43 & 14-1386 & 47.1 & J1058+0133 & J1037--2934 & J1037--2934 \\
 & 2018 Sep 20 \tablenotemark{\scriptsize3} & 44 & 14-1385 & 47.1 & J1037-2934 & J1037--2934 & J0904--5735 \\
2016.1.00464.S \tablenotemark{\footnotesize c} & 2016 Dec 03 \tablenotemark{\scriptsize5-7} & 40 & 14-662 & 48.3 & J1058+0133 & J1037--2934 & J1037--2934 \\
 & 2016 Dec 05 \tablenotemark{\scriptsize5-7} & 46 & 15-648 & 48.3 & J1058+0133 & J1037--2934 & J1058+0133 \\
 & 2016 Dec 07 \tablenotemark{\scriptsize5-7} & 45 & 14-609 & 48.3 & J1058+0133 & J1037--2934 & J1037--2934 \\
 & 2016 Dec 07 \tablenotemark{\scriptsize5-7} & 45 & 15-648 & 48.3 & J1058+0133 & J1037--2934 & J1058+0133 \\
 & 2016 Dec 07 \tablenotemark{\scriptsize5-7} & 39 & 15-596 & 48.3 & J1058+0133 & J1037--2934 & J1037--2934 \\
 & 2016 Dec 10 \tablenotemark{\scriptsize5-7} & 46 & 15-648 & 48.3 & J1058+0133 & J1037--2934 & J1058+0133 \\
 & 2016 Dec 11 \tablenotemark{\scriptsize5-7} & 46 & 15-636 & 48.3 & J1058+0133 & J1037--2934 & J1037--2934 \\
\enddata
\tablenotetext{a, b, c}{\ \ \ \ \ \ The Principal Investigators are K. I. \"{O}berg, L. I. Cleeves, and C. Walsh, respectively.}
\tablenotetext{1-7}{\ \ \ \ Link the transitions from Table~\ref{tab:lines} to the observation in which they are observed.}
\end{deluxetable*}

\begin{deluxetable*}{lcccccccc}[t!]
\tablecaption{Observed Formaldehyde Transitions \label{tab:lines}}
\tablewidth{\textwidth}
\tabletypesize{\footnotesize}
\tablehead{
\colhead{Transition}  & \colhead{Log$_{10}[A_{ij}]$}    & \colhead{$E_u$}       & \colhead{Robust\tablenotemark{\footnotesize a}}    & \colhead{Beam}        & \colhead{Chan. rms\tablenotemark{\footnotesize b}}             & \colhead{Mom-0 rms\tablenotemark{\footnotesize c}}   & \colhead{Int. Flux dens.\tablenotemark{\footnotesize d}} \vspace{-2mm} \\
\colhead{}      & \colhead{[s$^{-1}$]}            & \colhead{[K]}     & \colhead{}                    & \colhead{[$''\times'', ^{\circ}$]}     & \colhead{[mJy beam$^{-1}$]} & \colhead{[mJy beam$^{-1}$ km s$^{-1}$]}   & \colhead{[mJy km s$^{-1}$]}
}
\startdata
3$_{03}$--2$_{0 2}$(p)\tablenotemark{\scriptsize2} & $-$3.55037 & 20.96 & 0.5 & 0.49 $\times$ 0.33, $-87.8$ & 1.46 & 0.49 & $283\pm4$\\
3$_{12}$--2$_{11 }$(o)\tablenotemark{\scriptsize1} & $-$3.55724 & 33.45 & 2.0 & 0.53 $\times$ 0.50, $\hspace{6.5pt}88.7$ & 2.65 & 0.98 & $402\pm7$\\
4$_{04}$--3$_{03 }$(p)\tablenotemark{\scriptsize3} & $-$3.16102 & 34.90 & 2.0 & 0.35 $\times$ 0.29, $\hspace{6.5pt}64.7$ & 1.62 & 0.41 & $519\pm5$\\
4$_{22}$--3$_{21 }$(p)\tablenotemark{\scriptsize5} & $-$3.27994 & 82.12 & 2.0 & 0.51 $\times$ 0.47, $-60.3$ & 0.93 & 0.36 & $62\pm3$\\
4$_{31}$--3$_{30 }$(o)\tablenotemark{\scriptsize6} & $-$3.51653 & 140.9 & 2.0 & 0.51 $\times$ 0.47, $-62.2$ & 0.93 & 0.35 & $24\pm4$\\
4$_{32}$--3$_{31 }$(o)\tablenotemark{\scriptsize7} & $-$3.51684 & 140.9 & 2.0 & 0.51 $\times$ 0.47, $-62.2$ & 0.93 & 0.36 & $22\pm4$\\
5$_{15}$--4$_{14 }$(o)\tablenotemark{\scriptsize4} & $-$2.92013 & 62.45 & 2.0 & 0.35 $\times$ 0.28, $\hspace{6.5pt}83.5$ & 2.67 & 0.60 & $1118\pm7$\\
\enddata
\tablecomments{The rest frequency, Einstein $A$ coefficient, and upper state energy are taken from the LAMDA database.}
\tablenotetext{a}{The robust parameter used for Briggs weighting in the \textsc{CLEAN} process.}
\tablenotetext{b}{The channel rms is given at a common spectral resolution of 0.25 km~s$^{-1}$.}
\tablenotetext{c}{The moment-zero rms is determined through the bootstrapping described in Section~\ref{sec:obs}.}
\tablenotetext{d}{The integrated flux density is retrieved through summation of the emission retrieved by Keplerian masking of the emission cube.}
\tablenotetext{1-7}{\ \ \ \ Link the observations from Table~\ref{tab:obs_data} to the transitions.}
\end{deluxetable*}

\begin{figure*}[ht!]
    \includegraphics[width=\textwidth]{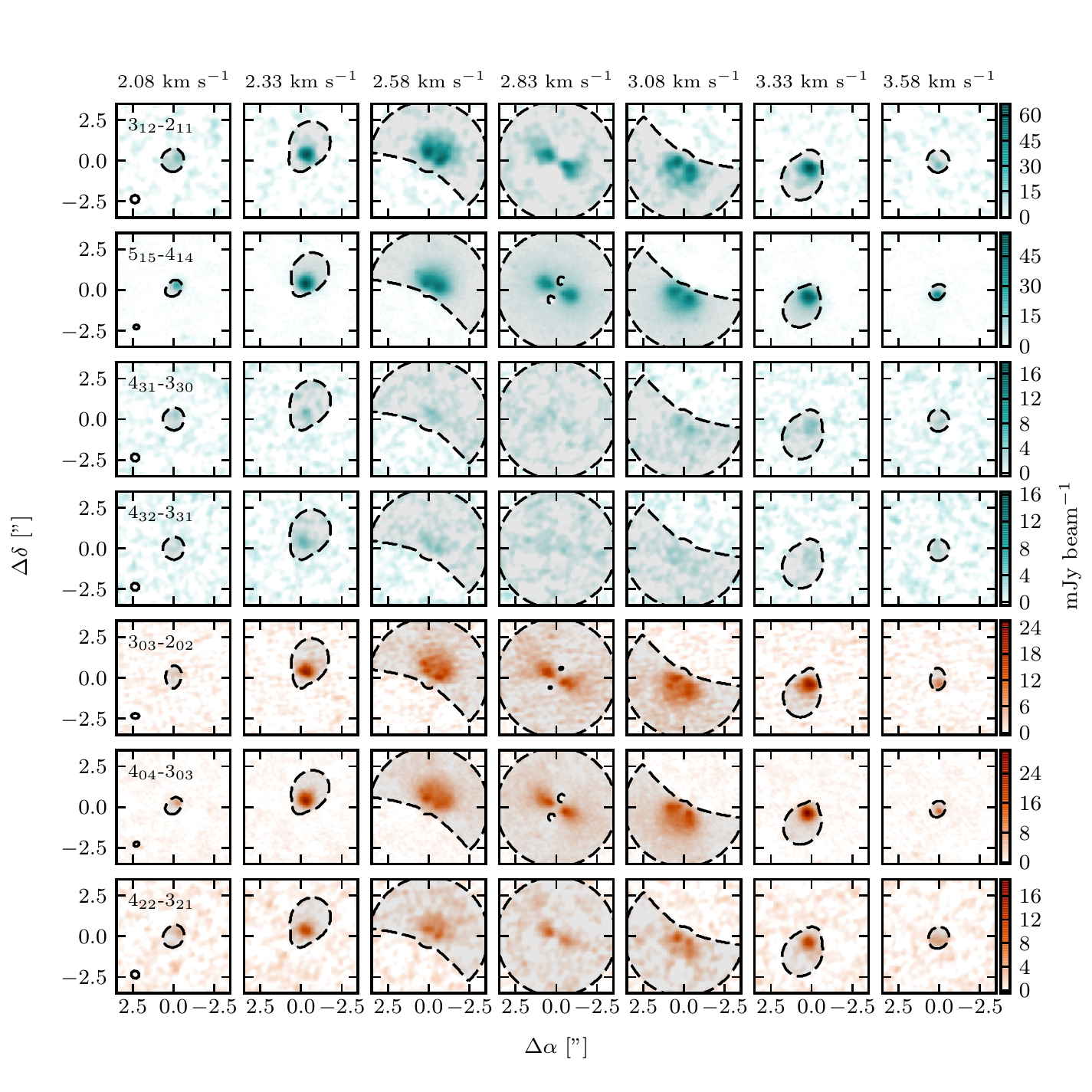}
\caption{Channel maps from the observed transitions at native spatial resolution with the Keplerian masked overlayed. The channel velocities are labelled on top in the V$_{\mathrm{LSR}}$ reference frame. The teal and bronze color correspond to the ortho- and para-spin isomer, respectively. Beam sizes are indicated by the ellipse in the left-bottom corner of the first column.}
\label{fig:channels}
\end{figure*}

\begin{figure*}[ht!]
    \includegraphics[width=0.96\textwidth]{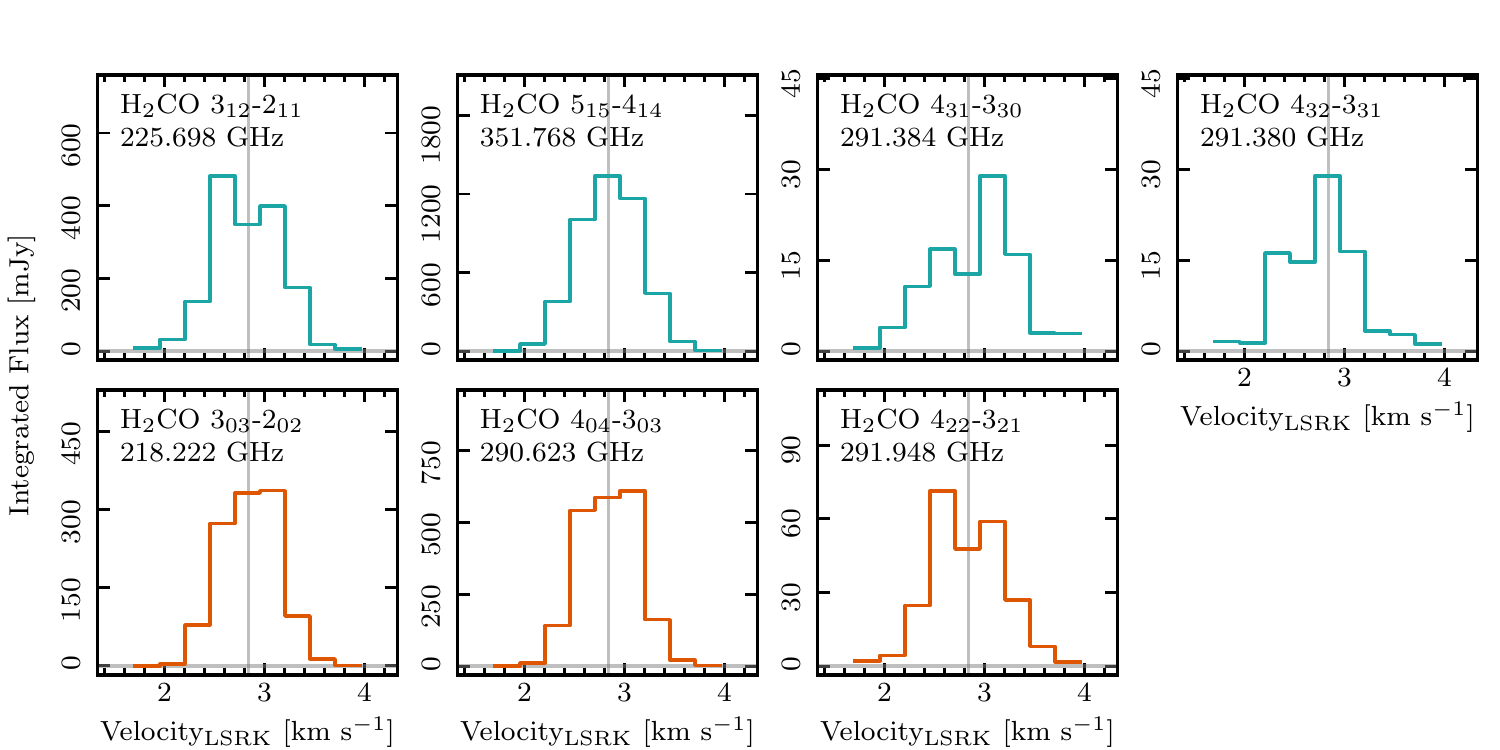}
\caption{Integrated spectra of the seven observed Keplerian masked \ch{H2CO} transitions, i.e. pixels are masked according to a model predicting the Keplerian rotation of TW~Hya. Top row, from left to right: o-\ch{H2CO} 3$_{12}$-2$_{11 }$, 5$_{15}$-4$_{14 }$, 4$_{31}$-3$_{30 }$, and 4$_{32}$-3$_{31 }$. Bottom row, from left to right: p-\ch{H2CO} 3$_{03}$-2$_{02 }$, 4$_{04}$-3$_{03 }$, and 4$_{22}$-3$_{21 }$. The vertical line indicates the systematic velocity of TW~Hya. \label{fig:int_spectra}}
\end{figure*}

The first protoplanetary disks in which \ch{H2CO} was detected are those around DM~Tau and GG~Tau \citep{1997_Dutrey_H2CO_DM/GG_Tau_A&A...317L..55D}. These detections were followed by the detection of \ch{H2CO} in LkCa~15 \citep{2003_Aikawa_H2CO_LkCa15_PASJ...55...11A,2004_Thi_H2CO_LkCa15_A&A...425..955T}. Although ground-breaking, the detections were only in the best case marginally spatially resolved and comparison to models \citep[e.g.][]{2003_Zadelhoff_UV_model_A&A...397..789V} to disentangle the origin of \ch{H2CO} was not feasible. 

Recent high-resolution observations with the Atacama Large Millimeter/Submillimeter Array (ALMA) of \ch{H2CO} transitions in 20 protoplanetary disks suggests that both gas-phase and solid-state formation of formaldehyde occurs, with their relative contributions varying across different disks \citep{2014_Marel_OphIRS48_A&A...563A.113V, 2015_Loomis_DMTAU_ApJ...809L..25L, 2017_Oberg_H2CO_ApJ...839...43O, 2017_Carney_H2CO_A&A...605A..21C, 2018_Kastner_V4046Sgr_ApJ...863..106K, 2018_Guzman_H2CO_ApJ...864..170G, 2019_Podio_DG_Tau_A&A...623L...6P, 2020_Pegues_H2CO_ApJ...890..142P, 2020_Garufi_DG_Tau_B_A&A...636A..65G}. Parametric model fits to resolved observations of \ch{H2CO} 3$_{12}$-2$_{11 }$ and 5$_{15}$-4$_{14 }$ in the disk of T~Tauri star TW~Hya find both warm and cold \ch{H2CO} components in compact and extended regions, respectively \citep{2017_Oberg_H2CO_ApJ...839...43O}. These studies have demonstrated that observations of multiple transitions allow for an improved determination of the rotational temperature and column density, which provides further constraints on the radial {\em and} vertical location of the emitting molecules and their origin. For example, in the disk of the Herbig~Ae star HD~163296, \citet{2018_Guzman_H2CO_ApJ...864..170G} derive for the first time a disk-averaged column density ratio of the ortho and para isomers of \ch{H2CO} in the range 1.8--2.8 with a rotational temperature of 24~K.

\begin{figure*}[ht!]
    \includegraphics[width=0.96\textwidth]{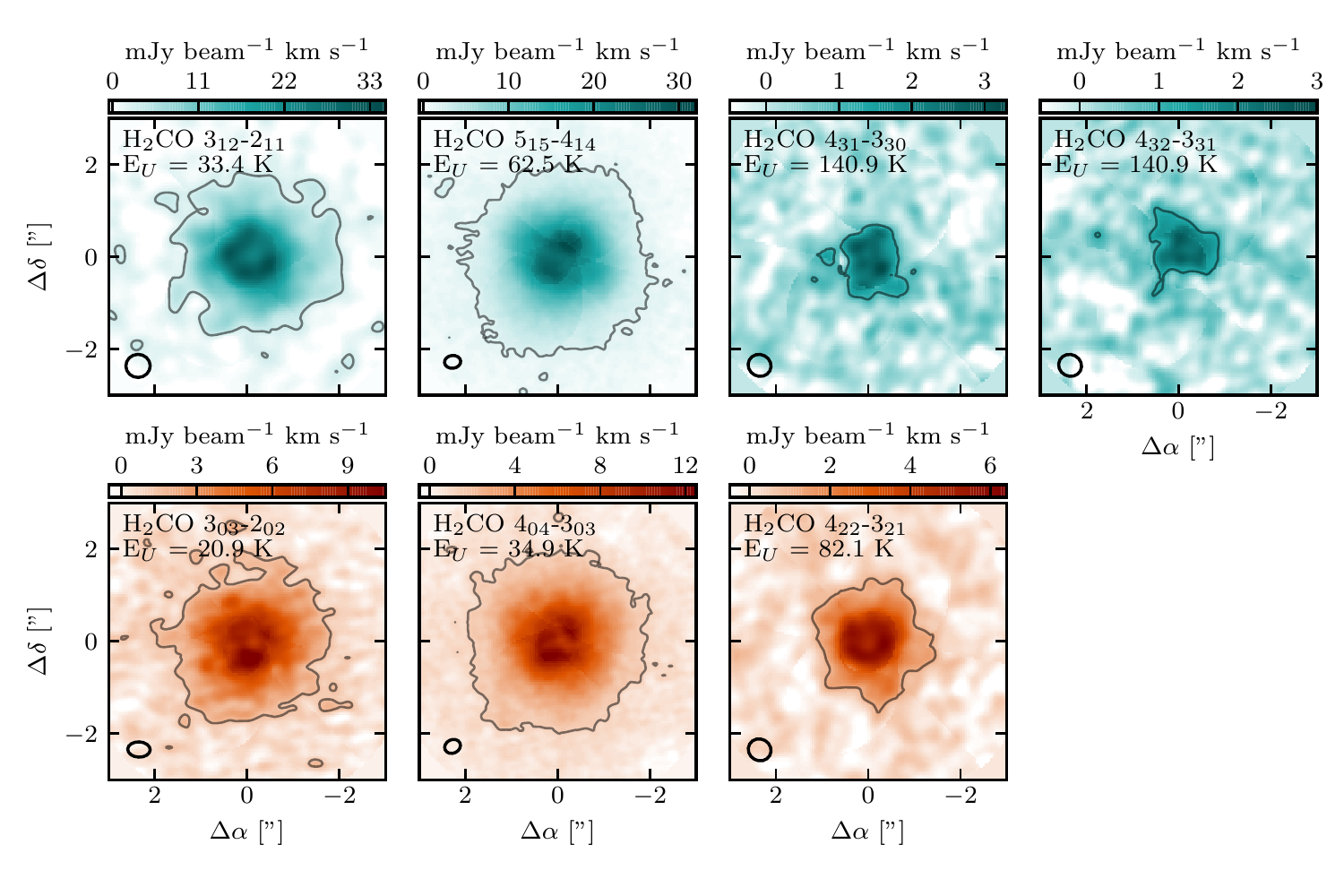}
\caption{Keplerian masked velocity integrated emission of observed \ch{H2CO} transitions at native spatial resolution ($V_{\rm LSR}=1.83$--$3.83$ km s$^{-1}$). Top row: ortho-spin isomer transitions; bottom row: para-spin isomer transitions. The contour in each panel depicts $3\sigma$, where $\sigma$ for each transition is taken from Table~\ref{tab:lines}. Beam sizes are indicated by the ellipse in the left-bottom corner of each panel. \label{fig:mom0}}
\end{figure*}

As first proposed by \citet{1984_Kahane_H2CO_A&A...137..211K}, the ortho-to-para ratio (OPR) of \ch{H2CO} could additionally shed light on the formation origins of this molecule. For example, \ch{H2CO} formed in warm gas would thermalize at the statistically expected OPR of 3.0, while cold formation, such as the CO ice hydrogenation pathways, would equilibrate the OPR to a lower value consistent with the grain temperature. The expectation is that the OPR is conserved from the moment of formation, since radiative transitions between ortho and para \ch{H2CO} are strictly forbidden. However, recent experimental work by \citet{2018_Hama_OPR_H2OApJ...857L..13H} shows that for \emph{water}, desorption {\em resets} the OPR to 3.0. If this is the case for \ch{H2CO}, other explanations for the observed low \ch{H2CO} OPR values are necessary and a cold-grain formation route cannot be inferred.

The disk around TW~Hya is an ideal laboratory to study the chemical origin of formaldehyde in detail. TW~Hya is the closest Sun-like star surrounded by a gas-rich protoplanetary disk, with a distance of 60.1~pc \citep{2018_GAIA_A&A...616A...1G}. Its disk has been studied extensively, in millimeter continuum and near-infrared scattered light, in various molecules including CO and isotopologues, and in a variety of chemical tracers \citep[e.g.][]{2012_Andrews_TW_Hya_ApJ...744..162A, 2015_Akiyama_TW_Hya_ApJ...802L..17A, 2016_Andrews_TW_Hya_ApJ...820L..40A, 2016_Walsh_CH3OH_ApJ...823L..10W, 2017_Oberg_H2CO_ApJ...839...43O, 2018_Huang_CODust_ApJ...852..122H, 2018_Teague_CS_ApJ...864..133T}. Spatially resolved observations of two \ch{H2CO} lines, 3$_{12}$-2$_{11 }$ ($0\farcs45 \times 0\farcs45$) and 5$_{15}$-4$_{14 }$ ($0\farcs47 \times 0\farcs41$), in the TW~Hya disk by \citet{2017_Oberg_H2CO_ApJ...839...43O} suggested that gas-phase formation dominates in the inner regions of the disk ($<$10 au) while grain-surface formation contributes beyond 15 au. 

In the current paper, we use a comprehensive multiline data set, including a wider range of upper state energies, 21--141~K, and now in both ortho- and para-spin-isomers, taken with ALMA toward the TW~Hya disk. These data allow us to \emph{directly} infer the radial and vertical structure of \ch{H2CO}, without having to rely on parametric models like those used by \citet{2017_Oberg_H2CO_ApJ...839...43O}. We aim to elucidate the formation of this key simple organic. Our data were obtained as part of an ALMA study (`TW Hya as a Chemical Rosetta Stone', PI L.I. Cleeves) aimed at a deep understanding of this object's chemistry, and, by extension, of that of other gas-rich protoplanetary disks. In this paper, observations of \ch{H2CO} from this ALMA project, together with archival ALMA data, are presented and used to explore the rotational temperature, column density, and ortho-to-para ratio of \ch{H2CO} in TW~Hya.  Section~\ref{sec:obs} describes the observational details and data reduction, \S\ref{sec:results} describes the resulting radial emission and excitation profiles, and \S\ref{sec:discuss} discusses the implications for the chemical origin of \ch{H2CO} across the TW~Hya disk.  Section~\ref{sec:sum} summarizes the main findings.

\section{Observations and Reduction} \label{sec:obs}

The data presented here were obtained as part of the ALMA project `TW~Hya as a chemical Rosetta stone' (2016.1.00311.S, PI Cleeves); additional, archival \ch{H2CO} data were taken from ALMA projects 2013.1.00114.S \citep{2017_Oberg_H2CO_ApJ...839...43O} and 2016.1.00464.S. Observational details (number of antennas, baseline ranges, on-source time and calibrators) are summarized in Table~\ref{tab:obs_data}. All data sets were processed through the standard ALMA calibration pipeline, after which self-calibration was applied. Data from 2013.1.00114.S is phase and amplitude self-calibrated on the continuum in the \ch{H2CO} spectral window using \textsc{CASA} 4.5 with timescales of 10--30 s. This improved the signal-to-noise ratio of the emission by a factor of $\approx$ 3. Phase self-calibration is applied to the data from 2016.1.00311.S using line free portions of the continuum. The solution interval is set to 30 seconds and polarization is averaged. Furthermore, the spectral windows are separately calibrated with a minimum signal-to-noise of 3 and minimum of 6 baselines per antenna. Data from 2016.1.00464.S is phase and amplitude self-calibrated with \textsc{CASA} 4.7.2 with two rounds of phase calibration, one over 30 second intervals and one over the integration time, and a single round of amplitude calibration. The signal-to-noise ratio in a CLEANed continuum image improved by a factor of $\approx$ 20. The final calibration tables are applied to the line-containing spectral windows.

Subsequent data processing was performed with \textsc{CASA} 5.6.1 \citep{2007_McMullin_CASA_ASPC..376..127M}. The continuum is subtracted using the \textsc{uvcontsub} task. Image reconstruction was performed with the \textsc{tclean} algorithm using the multiscale deconvolver \citep{1974_Hogbom_CLEAN_A&AS...15..417H, 2008_Cornwell_MS_ISTSP...2..793C} to reduce side lobes and increase the signal-to-noise ratio. Scales of $0''$, $0{\farcs}5$, $1{\farcs}0$, $2{\farcs}5$ and $5{\farcs}0$ were used for the multiscale deconvolver. No masking is applied as no significant difference was observed between the images with and without masking. Furthermore, masking creates a bias as scales used by the multiscale deconvoler larger then the masks are ignored by CLEAN. For the imaging of \ch{H2CO} 3$_{03}$-2$_{02}$ Briggs weighting with a robust parameter of 0.5 was used, resulting in a synthesized beam of $0{\farcs}49\times 0{\farcs}33$ and a good balance between the angular resolution and recovery of flux on all scales. All other \ch{H2CO} transitions were imaged with a robust parameter of 2.0, resulting in angular resolutions that closely match that of the 3$_{03}$-2$_{02}$ line and optimizing the sensitivity. The 3$_{12}$-2$_{11}$ transition is observed in only one execution block with a Maximum Recoverable Scale (MRS) of $2{\farcs}3$, less than the size of the disk line emission in several channels. Therefore, for this specific transition some ``smooth'' flux on larger scales may be missing with the exact amount depending on the details of the imaging reconstruction (for example, simple CLEAN vs multiscale CLEAN as applied by us). Given that the emission is not ``flat'' and is primarily peaked toward the star, missing flux is not expected to be a large contributor to the overall flux.

All images were CLEANed to three times the noise found by the \textsc{imstat} task in line free channels. The spectral resolution (channel width) of all cubes was set to 0.25 km~s$^{-1}$, close to the native resolution of the lowest resolution data (0.22 km~s$^{-1}$ for the 5$_{15}$-4$_{14}$ transition). For several of the observed transition our data combine observations from different ALMA configurations and require corrections to obtain correct flux densities and noise levels. This stems from disparities between the flux scale in Jy per synthesized beam of the (CLEAN-recovered) emission and Jy per dirty beam for the noise residuals (Loomis et al. 2020, in prep.). A final flux calibration uncertainty of 10\% is included in the further analysis as suggested by the ALMA Technical Handbook.

The CLEANed data cubes are masked according to the expected Keplerian rotation of TW~Hya. Pixels are masked on a per channel basis where no emission is expected to occur when the emitting gas in the protoplanetary disk around TW~Hya follows Keplerian rotation, \citep[e.g.,][]{2017_Salinas_DCO_A&A...606A.125S}. The mask is created with the disk parameters: PA $152^\circ$, inclination  $5^\circ$, and stellar mass 0.88 \(M_\odot\) from \citet{2018_Huang_CODust_ApJ...852..122H} with a systematic velocity of 2.83 km s$^{-1}$ ($V_{\rm LSR}$) and outer radius of 220~au, which corresponds to the edge of the gas disk as measured by CO \citep{2018_Huang_CODust_ApJ...852..122H}. Due to the nature of Keplerian masked moment-zero maps there is a nonuniform rms across the map, as described by \citet{2018_Bergner_CH3CN_ApJ...857...69B, 2020_Pegues_H2CO_ApJ...890..142P}. We follow these authors and bootstrap the uncertainty of the moment-zero maps and integrated flux densities by evaluating the rms of a large number of extractions across a similar number of randomly chosen line-free channels.

\section{Results} \label{sec:results}

\subsection{Observational Results} \label{sub:obs_res}
Emission in all seven targeted \ch{H2CO} transitions is clearly detected. Figure~\ref{fig:channels} shows the channel maps of the emission. Table~\ref{tab:lines} lists integrated flux densities of each transition extracted using Keplerian masking on the emission cubes. Values range from $1118\pm7$~mJy~km~s$^{-1}$ for the o-\ch{H2CO} 5$_{15}$-4$_{14 }$ line to $22\pm4$~mJy~km~s$^{-1}$ for the o-\ch{H2CO} 4$_{32}$-3$_{31 }$ line. Figure~~\ref{fig:int_spectra} shows the spectra integrated over the disk after Kepler masking.

\begin{figure}[t!]
    \includegraphics[width=0.48\textwidth]{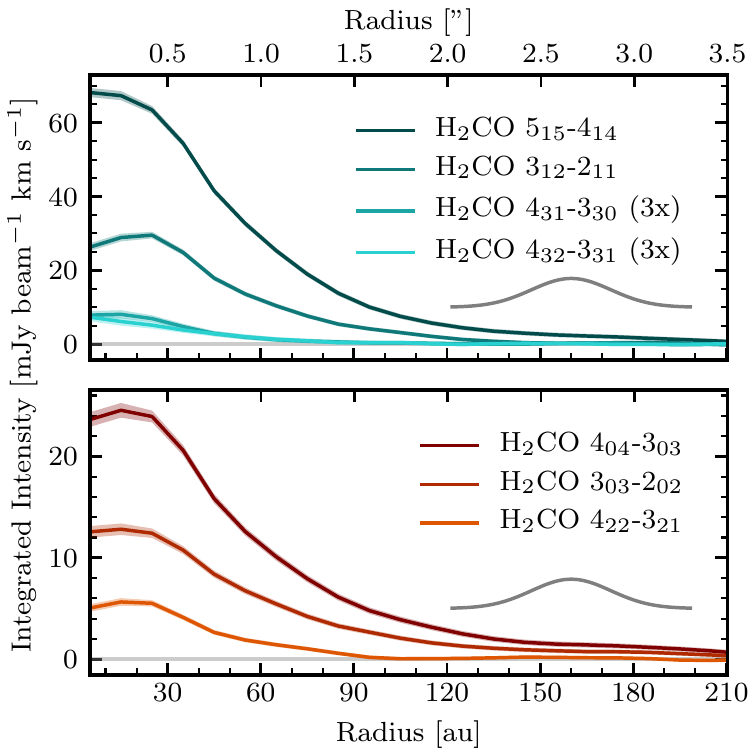}
\caption{Radial intensity profiles of the observed ortho (top panel) and para (bottom panel) \ch{H2CO} transitions retrieved from the Keplerian masked moment-zero maps. The displayed uncertainties do not contain the $10\%$ flux calibration error. All data have a common spatial resolution of 30~au depicted by the Gaussian in the bottom right.}
\label{fig:rad_prof}
\end{figure}

The channel maps clearly show that the emission follows the velocity pattern of a disk in Keplerian rotation. Using the expected region of emission in each velocity channel integrated intensity (zero moment) maps are obtained and shown in Fig.~\ref{fig:mom0}. These images show that the \ch{H2CO} emission is concentrated in a ring with a radius of $0{\farcs}3$ (18~au), with a broad fainter brim of emission as was also seen by \citet{2017_Oberg_H2CO_ApJ...839...43O}. From these Keplerian masked integrated intensity images, radial emission profiles are extracted by annular averaging in 10~au wide bins (Fig.~\ref{fig:rad_prof}). Uncertainty levels of the radially averaged intensities are calculated by dividing the moment-zero rms with the square-root of the number of independent beams present in that bin. To bring all data on the same angular resolution of $0{\farcs}5$, the \ch{H2CO} 3$_{03}$-2$_{02 }$, 4$_{04}$-3$_{03 }$, and 5$_{15}$-4$_{14 }$ were re-imaged using CLEAN and respective \textsc{uvtaper} of [$0{\farcs}0, 0{\farcs}35, -87.8^{\circ}$], [$0{\farcs}23, 0{\farcs}33, 64.7^{\circ}$], and [$0{\farcs}18, 0{\farcs}30, 83.5^{\circ}$] before the radial intensity profiles of these transitions were extracted. It should be noted that the 4$_{31}$--3$_{30 }$ and 4$_{32}$--3$_{31 }$ transitions have very similar excitation parameters and are thus difficult to distinguish in Figure~\ref{fig:rad_prof}.

\subsection{\ch{H2CO} Excitation Temperature and Column Density} \label{sub:excit}

\subsubsection{Rotational diagram analysis}
The wide range of upper-state energies of 21--141~K of the detected \ch{H2CO} lines allow for well constrained estimates of the excitation temperatures and column densities of the ortho and para isomers through a rotation diagram analysis \citep[e.g.,][]{1999_Goldsmith_RTD_ApJ...517..209G}. Since the gas densities in the disk \citep[as estimated from the models of][]{2015_Cleeves_TW_Hya_ApJ...799..204C,2016_Kama_DALI_A&A...592A..83K} typically exceed the critical density of the targeted transitions, $n_{\rm H_2} \sim 10^6$--$10^7$ cm$^{-3}$, the molecule's excitation is likely in local thermal equilibrium (LTE), even in the outer region of the disk. Derived excitation temperatures therefore are a reliable estimate of the kinetic temperature of the emitting gas, provided that the emission is optically thin. In the treatment outlined below, specific allowance is made for moderately optically thick emission.

\begin{figure}[t!]
    \includegraphics[width=0.48\textwidth]{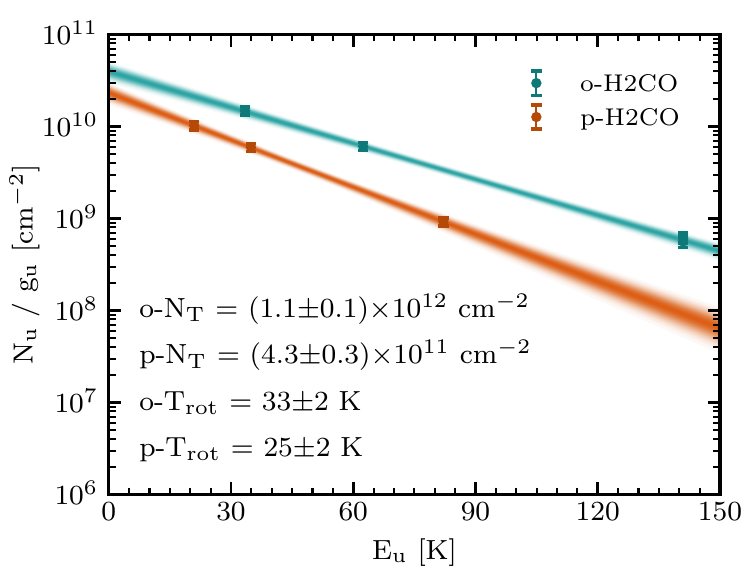}
\caption{Rotational diagram of disk-averaged \ch{H2CO} flux density values from Table~\ref{tab:lines}. The teal and bronze color represent the level populations of the ortho- and para-spin isomers, respectively. The markers show the data and the lines show random draws from the posterior distribution retrieved by the \textsc{emcee} fitting procedure. \label{fig:disk_avg_mcmc}}
\end{figure}

The line intensity, $I_{\nu}$, follows from the column density of the upper-state level in the optically thin limit, $N^{\rm thin}$, as
\begin{equation} \label{eq:I_v}
    I_\nu = \frac{A_{ul}N_u^{\rm thin}hc}{4\pi\Delta v},
\end{equation}
where $A_{ul}$ is the Einstein $A$ coefficient and $\Delta v$ the velocity width of the emission line. Rewriting equation~\ref{eq:I_v} and substituting the source brightness $I_{\nu}$ as flux per solid angle, $S_{\nu} / \Omega$, gives
\begin{equation} \label{eq:N_u^thin}
    N^{\rm thin}_{u} = \frac{4\pi S_\nu \Delta v}{A_{ul} \Omega hc}.
\end{equation}
Here, $S_\nu$ is the flux density extracted from the integrated spectra or the radial flux profiles, and $\Omega$ is the total solid angle from which the emission is extracted. If the emission is not fully optically thin, the column density of the upper-state level, $N_u$, follows from the optically thin limit by applying a correction for line optical depth,
\begin{equation} \label{eq:N_u}
    N_u = N_u^{\rm thin}\,\frac{\tau}{1-e^{-\tau}}
\end{equation}
where $\tau$ is the optical depth at the center of the line. This line opacity $\tau$ is given by
\begin{equation} \label{eq:tau}
    \tau = \frac{A_{ul} N_u^{\rm thin} c^3}{8\pi \nu^3 \Delta v}(e^{h\nu/kT_{\rm rot}}-1)
\end{equation}
where $\nu$ is the rest-frequency of the transition. An upper limit to the opacity follows from assuming a line width $\Delta v$ equal to the disk-averaged FWHM of the intrinsic line, estimated to be 0.275 km~s$^{-1}$. This value is estimated from the FWHM of the Keplerian corrected integrated spectra acquired with \textsc{GoFish} \citep{GoFish}. Finally, the total column density, $N_{\rm tot}$, is related to the upper-state level populations through the Boltzmann equation,
\begin{equation} \label{eq:Nu/gu}
    \frac{N_u}{g_u} = \frac{N_{\rm tot}}{Q(T_{\rm rot})} e^{-E_u/kT_{\rm rot}},
\end{equation}
where $g_u$ is the degeneracy of the corresponding upper state level, $Q$ the partition function of \ch{H2CO}, $E_u$ the upper state level energy, and $T_{\rm rot}$ the rotational temperature of \ch{H2CO}. The upper state degeneracy, upper state energy, Einstein $A$ coefficient, and frequency of each transition are extracted from the Leiden Atomic and Molecular Database (LAMDA) \citep{2005_Schoier_LAMDA_A&A...432..369S}. The partition function for \ch{H2CO} is constructed from the rotational ground states taken from the ExoMol database \citep{2015_Refaie_ZH2CO_MNRAS.448.1704A, 2020_ExoMol_Atoms...8....7W}. In order to independently investigate the nuclear spin isomers a separate partition function is created for each of the isomers. The ExoMol database assumes an OPR of three. This OPR is incorporated in their state degeneracies and is removed by dividing by three to match the state degeneracies of the LAMDA database. The partition function is constructed by summing over the possible internal \ch{H2CO} ground states,
\begin{equation} \label{eq:part}
    Q(T_{\rm rot}) = \sum_{i} g_i e^{-E_i/kT}
\end{equation}
where $g_i$ is the degeneracy and $E_i$ the energy of state $i$. These separate partition functions allow independent determination of the column densities of each spin isomer. As is customary for rotation diagram analyses, the column density $N_{\rm tot}$ and rotation temperature $T_{\rm rot}$ are retrieved from the intercept and slope, respectively, of $\ln(N_u/g_u)$ vs $E_u$ (Fig.~\ref{fig:disk_avg_mcmc}). Following \citet{2018_Loomis_CH3CN_ApJ...859..131L} and \citet{2018_Teague_CS_ApJ...864..133T}, we create a likelihood function from Equation~\ref{eq:Nu/gu} and use \textsc{emcee} \citep{2013_Foreman_MCMC_PASP..125..306F} to retrieve posterior distributions for $T_{\rm rot}$ and $N_{\rm tot}$ from the observed \ch{H2CO} transitions. This rotational diagram fitting procedure is applied to each of the spin isomers separately.

\subsubsection{Disk-averaged rotational diagram}
The disk integrated flux densities of Table~\ref{tab:lines}, analysing each spin isomer separately, yields disk averaged column densities and rotational temperatures of $(1.1\pm0.1)\times10^{12}$~cm$^{-2}$, $33\pm2$~K and $(4.3\pm0.3)\times10^{11}$~cm$^{-2}$, $25\pm2$~K for ortho- and para-\ch{H2CO}, respectively. These values result in a disk-averaged OPR of $2.49\pm0.23$. Uncertainties are the 16th and 84th percentiles of the posterior distributions, corresponding to 1~sigma. The disk-averaged line opacities range from 0.002 to 0.049, confirming the assumption of optically thin emission. However, one should note that this assumes a uniform distribution of \ch{H2CO} across the entire disk which is not the case, as seen in Figure~\ref{fig:rad_prof}. The opacities will be larger in the inner region where column densities are higher, as shown in Section~\ref{sub:rad_res_rot}.

\citet{2019_Carney_H2CO_CH3OH_A&A...623A.124C} investigated the disk-averaged ratio of \ch{CH3OH} with respect to \ch{H2CO} for the protoplanetary disks around HD~163296 and TW~Hya. Specifically for TW~Hya they found a \ch{CH3OH/H2CO} ratio of 1.27$\pm$0.13. In their work the average \ch{H2CO} and \ch{CH3OH} column densities are derived self-consistently from the integrated line intensity of one transition and an assumed excitation temperature, \citep[see Eq (1)][]{2019_Carney_H2CO_CH3OH_A&A...623A.124C}. The \ch{H2CO} column density is found to be $3.7\times10^{12}$~cm$^{-2}$, which is approximately 2.4 times higher than the derived average total \ch{H2CO} column density of $(1.5\pm0.1)\times10^{12}$~cm$^{-2}$ in this work. The lower total \ch{H2CO} column density derived through rotational diagram analysis pushes the \ch{CH3OH/H2CO} ratio up to a value of 3.1$\pm$0.4. However, it should be noted that the \ch{CH3OH} column density used in this work is taken from \citet{2019_Carney_H2CO_CH3OH_A&A...623A.124C} and is thus not derived self consistent with the \ch{H2CO} column density.

The rotational temperatures of both spin isomers are not identical, with a slightly higher value of 33~K found for ortho-\ch{H2CO} compared to 26~K for para-\ch{H2CO}. If the OPR is in thermal equilibrium, such a difference is expected: the ortho isomer is the more abundant in warmer gas compared to the para isomer, resulting in a higher rotational temperature for the former when averaging its emission over the disk. However, there may also be a systematic bias, because the detected o-\ch{H2CO} lines extend over a larger range of upper-level energies (up to 141~K) compared to the p-\ch{H2CO} lines (up to 82~K), thus naturally probing higher excitation gas. Section~\ref{sub:rad_res_rot} explores further explanations, folding in spatially resolved information.

\subsubsection{Radially resolved rotational diagram} \label{sub:rad_res_rot}

\begin{figure}[t!]
    \includegraphics[width=0.48\textwidth]{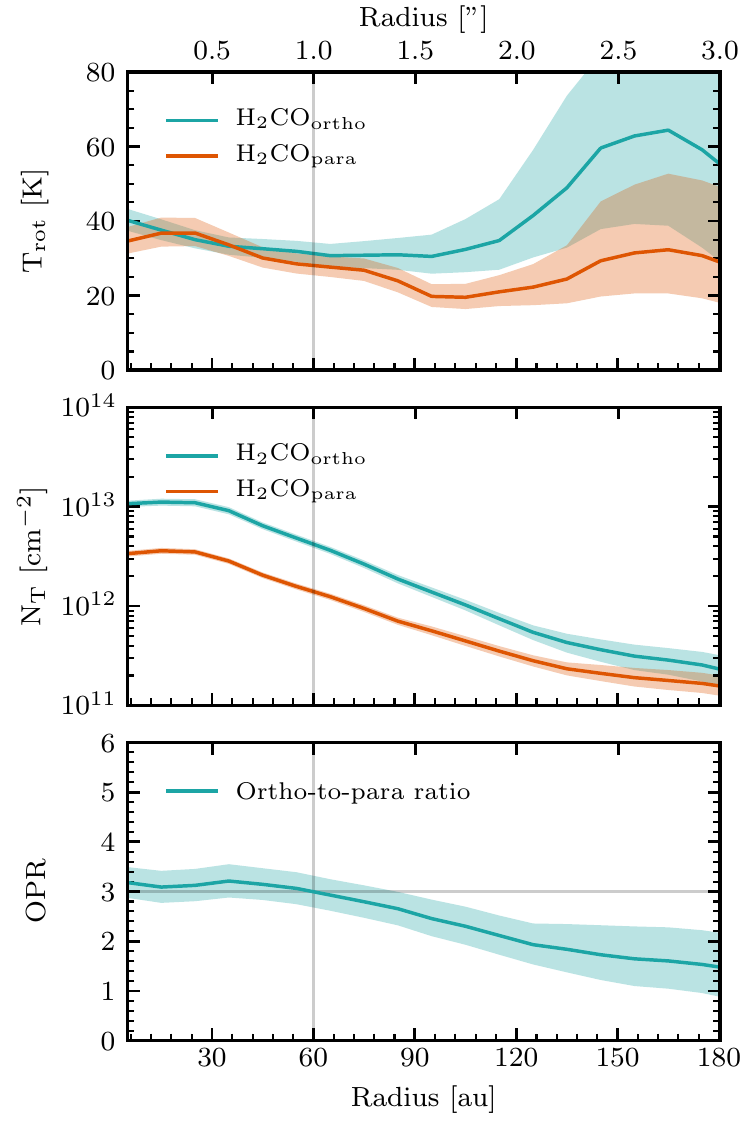}
\caption{Radially resolved \ch{H2CO} temperature (top panel), column densities (middle panel), and OPR (bottom panel) of \ch{H2CO} obtained from our rotational diagram analysis using the radial emission profiles of Fig.~\ref{fig:rad_prof}. Teal and bronze colors represent the ortho- and para-spin isomers, respectively. Shaded areas depict 1$\sigma$ uncertainties. The gray vertical and horizontal line depict the mm-dust continuum edge and the high temperature OPR limit of 3.0, respectively. \label{fig:MCMC_rad}}
\end{figure}

\begin{figure}[t!]
    \includegraphics[width=0.48\textwidth]{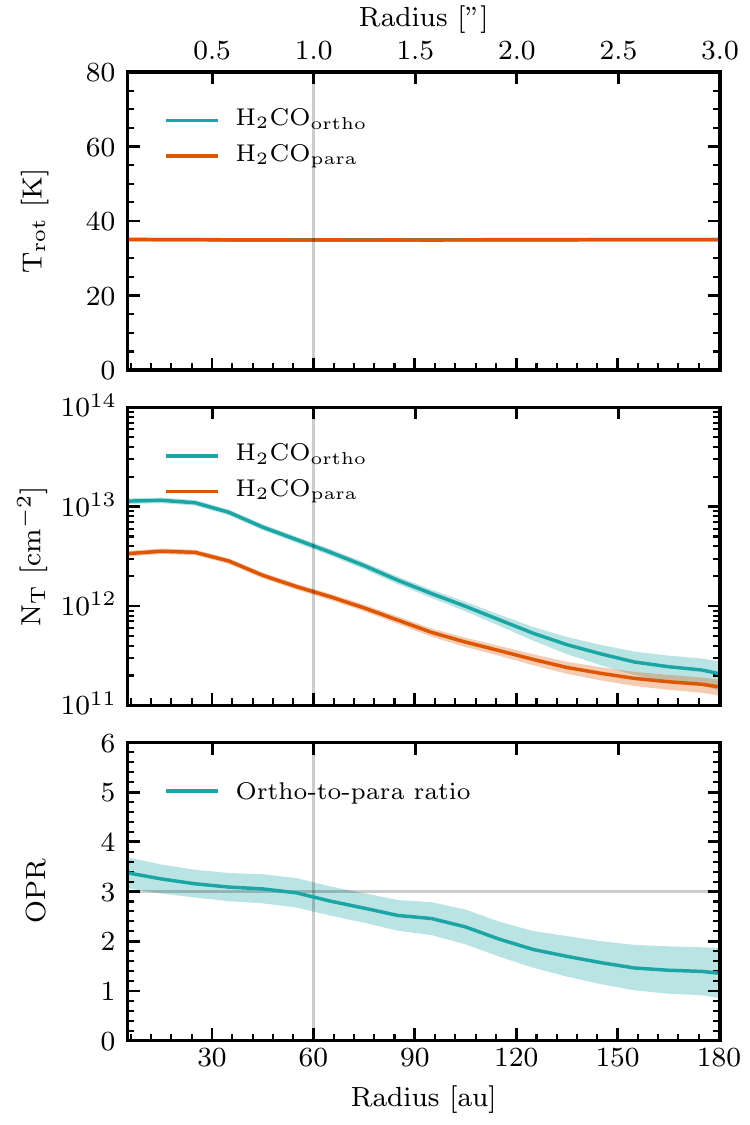}
\caption{Radially resolved \ch{H2CO} temperature (top panel), column densities (middle panel), and OPR (bottom panel) of \ch{H2CO} obtained from our rotational diagram analysis with a fixed rotational temperature at 35~K. Teal and bronze colors represent the ortho- and para-spin isomers, respectively. Shaded areas depict 1$\sigma$ uncertainties. The gray vertical and horizontal line depict the mm-dust continuum edge and the high temperature OPR limit of 3.0, respectively. \label{fig:MCMC_rad_fixed}}
\end{figure}

The same rotational diagram analysis as carried out above for the disk-integrated fluxes, can also be performed as function of radius, using the Keplerian masked moment-zero images, all restored to a common resolution of $0{\farcs}5$, and the corresponding radial intensity profiles of Fig.~\ref{fig:rad_prof}. As depicted in Fig.~\ref{fig:MCMC_rad}, the column densities in the $0{\farcs}5$ beam peak at $\sim$20~au with values of $(1.1\pm0.1)\times10^{13}$~cm$^{-2}$ and $(3.6\pm0.3)\times10^{12}$~cm$^{-2}$ for o-\ch{H2CO} and p-\ch{H2CO}, respectively. The opacities are largest in the inner region where column densities are highest. The largest optical depths are found to be $\tau=0.47$ and $\tau=0.35$ for the 5$_{15}$-4$_{14 }$ and 3$_{12}$-2$_{11 }$ transition, respectively. Beyond 85~au all the opacities drop to a value of $\tau < 0.1$. Therefore, all transitions only require a moderate correction for opacity. 

The rotational temperatures of both the ortho- and para-\ch{H2CO} are found to be consistent with each other within 80~au. The averaged rotational temperatures drops as function of radius from $37\pm5$~K at the center of the first radial bin (5~au), to $29\pm5$~K at 75~au. These observed rotational temperature are below the freeze out temperature of gas-phase \ch{H2CO} at approximately 80~K \citep{2012_Noble_H2CO_TPD_A&A...543A...5N, 2020_Pegues_H2CO_ApJ...890..142P}. Beyond 80~au the rotational temperatures deviate from each other with a two sigma tension. This explains the observed difference in rotational temperatures from the disk-averaged analysis. At radii beyond 120~au no meaningful constraints on the temperature are found due to the low signal-to-noise ratio in the majority of the transitions.

The observed temperature difference between the ortho- and para-spin isomer from the disk-averaged analysis can be traced back to two transitions: 3$_{12}$--2$_{11 }$ and 4$_{22}$--3$_{21 }$. The 3$_{12}$--2$_{11 }$ transition is observed with a MRS of $2{\farcs}3$. This transition has an upper-state energy of 33~K, the lowest of the ortho-spin isomers in this data set. With potentially missing extended emission on scales of $2{\farcs}3$ and larger, the fit may result in higher rotational temperatures on these scales. The 4$_{22}$--3$_{21 }$ transition has an upper-state energy of 82~K, which is the highest of the para-spin isomers in this data set. Combined with the drop in emission of the radial profile from 90 to 140~au, this may result in lower rotational temperatures in this region. Additional observations with a more robust coverage of the extended emission are needed to determine the \ch{H2CO} emitting temperature more robustly.

Dividing the column density profiles of o-\ch{H2CO} and p-\ch{H2CO} yields the radially resolved OPR, which is consistent with an OPR of 3.0, the high temperature limit, in the inner 60~au and drops to lower values at larger radii, e.g. $2.0\pm0.5$ at a distance of 120~au. We rule out that the  temperature bias described in the previous paragraph has an impact on the derived OPR. Repeating the analysis at fixed rotational temperatures ranging from 30 to 40 K, and omitting the 3$_{12}$--2$_{11}$ transition that may miss extended emission, we recover the same downward OPR trend. The fixed rotational temperature model of 35~K is shown in Fig.~\ref{fig:MCMC_rad_fixed}. We therefore conclude that the radial decrease in the OPR beyond the mm-dust continuum is robust.

\section{Discussion} \label{sec:discuss}

\subsection{The inner \ch{H2CO} line emission decrease}
Our resolved, multi-line observations of H$_2$CO in TW~Hya broadly indicate a flattening or decrease in flux interior to  20 au. An inner deficit in the intensity profile has been seen in other disks and other molecular lines, and can be attributed either to real decreases in column density or to line or continuum opacity \citep{2012_Andrews_TW_Hya_ApJ...744..162A,2016_Cleeves_IM_Lup_ApJ...832..110C,2016_Isella_HD_163296_PhRvL.117y1101I,2017_Loomis_AA_Tau_ApJ...840...23L}. Continuum over-subtraction due to optically thick lines can be ruled out in this case, given the consistently low optical line depths ($<0.5$, see \S\ref{sub:excit}). Continuum opacity is also likely not the only explanation, because emission of other molecules at similar wavelengths are centrally peaked (e.g., C$^{18}$O $J = 3-2$ imaged by \citet{2016_Schwarz_CO_ApJ...823...91S}). However, it should be noted that \citet{2018_Huang_CODust_ApJ...852..122H} find the millimeter wavelength spectral index inside  $<$ 20~au to be 2.0, indicative of optically thick continuum emission. Depending on the height where the line emission originates, \emph{some} of the drop in emission may be due to an optically thick continuum. Nonetheless, our data show evidence of \ch{H2CO} inside of 20~au, the approximate CO snow line location in this disk \citep{2016_Schwarz_CO_ApJ...823...91S,2017_Merel_N2H+_A&A...599A.101V,2017_Zhang_NatAs...1E.130Z}, indicative of active gas phase \ch{H2CO} chemistry as was also found in \citet{2017_Oberg_H2CO_ApJ...839...43O}.

\subsection{Gas-phase vs grain-surface formation of \ch{H2CO}}
In interstellar environments, \ch{H2CO} forms by a combination of CO ice hydrogenation and neutral-neutral gas-phase reactions, specifically CH$_3$ + O and CH$_2$ + OH \citep[e.g.,][]{2015_Loomis_DMTAU_ApJ...809L..25L}. For CO ice to exist in abundance, the dust grain temperature must be quite low, below 25~K depending on the binding surface. This temperature is much lower than the thermal desorption temperature of \ch{H2CO}, therefore \ch{H2CO} formed by this mechanism requires subsequent non-thermal desorption to produce observable gas-phase quantities.  Previous work by \citet{2015_Loomis_DMTAU_ApJ...809L..25L} and \citet{2017_Oberg_H2CO_ApJ...839...43O} found that a combination of gas-phase and solid-state chemistry likely contributes to the observed gaseous \ch{H2CO} in disks. In TW Hya specifically, \citet{2017_Oberg_H2CO_ApJ...839...43O} suggests that the \ch{H2CO} ring near the CO snow line could be evidence of a CO-ice regulated chemistry \citep[see also][]{2013_Qi_H2CO_N2H+_ApJ...765...34Q}. In addition, a number of sources, e.g. HD~163296, CI~Tau, DM~Tau, and AS~209 \citep[see][]{2020_Pegues_H2CO_ApJ...890..142P}, show an increase or ring in \ch{H2CO} in the outer disk. A clear example of a secondary increase is found in HD~163296 at a radius of $\sim$250~au \citep{2017_Carney_H2CO_A&A...605A..21C}. These authors suggest that an additional formation route related to CO ice may be opening up at this location, or that increased penetration of ultraviolet radiation boosts gas-phase formation of \ch{H2CO}. 

Previous work by \citet{2017_Oberg_H2CO_ApJ...839...43O} showed a similar emission bump in the 3$_{12}$--2$_{11 }$ transition near the mm-dust continuum edge at 60 au. Our imaging of the same data at similar spatial resolution does not show this emission bump. We attribute the difference to the Maximum Recoverable Scale of $2{\farcs}3$ of this data set in combination with the different applied CLEAN method (multiscale CLEAN), since this method is expected to yield a more reliable result for extended emission.

The near-constant rotational temperature of 30--40~K found here for \ch{H2CO} suggests that the emission arises from an elevated layer in TW Hya's disk, well above the CO snow surface. In models of the TW~Hya disk \citep[][Calahan et al. 2020, in prep.]{2013_Bergin_HD_Natur.493..644B,2015_Cleeves_TW_Hya_ApJ...799..204C,2016_Kama_DALI_A&A...592A..83K}, these temperatures are found at normalized heights of $z/R\ge 0.25$. Additionally, recent observations of the edge-on younger embedded disk IRAS~04302 also show that the bulk of the \ch{H2CO} emission arises from $z/R\sim$ 0.21 - 0.28 \citep{2020_Merel_emb_disks_arXiv200808106V, 2020_Podio_emb_disks_arXiv200812648P}. At these heights, sufficient UV can penetrate to induce photodesorption of \ch{H2CO} since the bulk of the small dust has grown and is very settled \citep[e.g.,][]{2004_Dullemond_Dust_set_A&A...421.1075D,2014_Testi_Dust_evo_prpl.conf..339T}. These same UV photons also induce efficient gas-phase formation of \ch{H2CO} by radical production, since the two radical-radical gas phase reactions that form \ch{H2CO} are barrierless. \citet{2020_Teague_CN_TW_Hya_arXiv200711906T} found CN at similar heights in TW~Hya, a molecule which is formed mainly through UV irradiation \citep{2018_Cazzoletti_CN_TW_Hya_A&A...609A..93C}. Our data therefore suggest that gas-phase formation is likely important to explain the observed gas-phase \ch{H2CO} across the entire disk of TW~Hya. 

Interestingly, \citet{2015_Loomis_DMTAU_ApJ...809L..25L} found that gas-phase chemistry alone underproduced the observed column density of \ch{H2CO} in the DM Tau protoplanetary disk. However, it is important to note that the modeling carried out in \citet{2015_Loomis_DMTAU_ApJ...809L..25L} either fully turned off CO-hydrogenation or left on the full CO-hydrogenation pathway up to forming CH$_3$OH.
While detailed chemical modeling is beyond the scope of the present paper, we examined the reaction rates from the existing \citet{2015_Cleeves_TW_Hya_ApJ...799..204C} TW Hya chemical model with a dust surface area reduction of 85\%. The latter is invoked to emulate the effects of dust settling and radial drift, which significantly reduce the effective solid surface for ice chemistry to occur \citep{2011_Hogerheijde_ice_h2o_Sci...334..338H,2018_Bergin_Cleeves_book_haex.bookE.137B}.
In the layer where \ch{H_2CO} is abundant ($z/R > 0.25$), the two gas phase pathways with O and OH are far more efficient than CO ice hydrogenation due to the warm temperatures of the surface layers. From these initial tests it also appears that, although \ch{H2CO} formed in the gas-phase is easily photodissociated, subsequent freeze out of the resulting HCO radicals reforms \ch{H2CO}, as the hydrogenation step involved is barrierless \citep{2009_Fuchs_hydro_A&A...505..629F}. Further modeling is needed to confirm this symbiotic gas-grain relationship in \ch{H2CO} formation. The key role of UV-induced gas-phase chemistry has been seen in other models. \citet{2014_Walsh_Chem_model_A&A...563A..33W} find that \ch{H2CO} can be efficiently formed through gas-phase chemistry alone around a typical T Tauri star. They find a fractional abundance with respect to $n_{\rm H}$ of $10^{-10}$ to $10^{-9}$, which translates in their models to column densities between $10^{12}$ and $10^{13}$~cm$^{-2}$, very similar to the values obtained from our observations. It should be noted that although gas-phase chemistry is sufficient to explain observed \textit{gas-phase} column densities it does not imply that \textit{solid-state} formation does not occur in the disk midplane. The chemical models generally produce 5 orders of magnitude more solid-state \ch{H2CO} in the disk midplane.

\subsection{Constraints from the \ch{H2CO} OPR on the formation}
The smooth radial \ch{H2CO} column density profile and near-constant excitation temperature are consistent with a single origin of the observed \ch{H2CO}, namely gas-phase formation. Is this consistent with the radial gradient in OPR that is also observed? As first proposed by \citet{1984_Kahane_H2CO_A&A...137..211K}, the OPR of \ch{H2CO} -- if distributed according to a Boltzmann distribution -- drops below 3.0 for spin temperatures below $\sim$35~K and reaches 2.0 for a spin temperature of $\sim$12~K \citep[cf.\ Fig.~10 of][]{1984_Kahane_H2CO_A&A...137..211K}. The spin temperature is thought to correspond to the formation temperature of the molecule since the gas-phase nuclear spin conversion time for non reactive collisions is longer than the \ch{H2CO} lifetime \citep{2006_Tudorie_NSC_H2CO_A&A...453..755T}. Within 60~au, the inferred \emph{rotational} temperatures of 30--40~K are consistent with the \emph{spin} temperatures of $\gtrsim 27$~K found from the OPR. Outside 60~au, and especially outside 120~au, the OPR suggest a spin temperature of 10--17~K while the (poorly constrained) rotational temperature exceeds 20~K ($1\sigma$).

A low OPR, and corresponding low formation temperature, has been invoked as evidence for formation of \ch{H2CO} in the ice, during the prestellar phase or in cold regions of the disk, and subsequent release in the gas. This is based on the expectation that the OPR is conserved from the moment of formation, because radiative transitions between ortho and para \ch{H2CO} are strictly forbidden. However, recent experimental work by \citet{2018_Hama_OPR_H2OApJ...857L..13H} shows that non-thermal desorption of para-enriched water ice at 11~K causes the OPR to revert to 3.0, as expected for higher temperatures. For water, this is explained by the fact that water molecules in the ice cannot rotate because of hydrogen bonds in the ice matrix. This restriction results in a quasi-degeneracy of the ortho- and para-\ch{H2O} states in the solid-state. Furthermore, theoretical studies on solid-state \ch{H2O} propose that rapid nuclear-spin conversion in the solid-state is possible through intermolecular proton-proton magnetic dipolar interactions \citep{2006_Limbach_NSC_CPC...V7...I3, 2008_Buntkowsky_NSC_ZPC...V222...I7}. Similar to \ch{H2O}, \ch{H2CO} will also be rotationally hindered in the solid-state, and an OPR of 3.0 may be expected on release into the gas-phase, even when formed at low temperatures. However, the extent of rotational hindrance of \ch{H2CO} in an apolar \ch{CO} matrix has to be investigated theoretically or experimentally before a conclusive statement can be made.

If we accept that the OPR reflects the temperature of the \ch{H2CO} formation in the ice, our observed values indicate that only outside 60 au does the observed \ch{H2CO} emission contain a contribution originating in the ice. Although not very well constrained, the rotational temperature at these radii exceeds the freeze-out temperature of CO, $<$ 21~K \citep{2016_Schwarz_CO_ApJ...823...91S}, suggesting that some vertical transport of \ch{H2CO} formed in the midplane through hydrogenation is required. Given the low turbulence in the TW~Hya disk \citep{2018_Flaherty_Turbu_TW_Hya_ApJ...856..117F}, it is not immediately clear what mechanism can efficiently explain this vertical transport. Alternatively, solid-state \ch{H2CO} can be inherited from the prestellar stage \citep[e.g.,][]{2011_Visser_inheritance_A&A...534A.132V}. This inherited \ch{H2CO} ice could then non-thermally desorb in an elevated layer in the protoplanetary disk stage before it settles to the disk midplane. If, however, the observed \ch{H2CO} has an inherited origin and we assume the OPR is preserved we would expect a constant value. The observed OPR ranging from 3.0 to 2.0 in this single monotonic component would thus require an external influence, e.g., different desorption conditions or subsequent disk gas-phase chemistry. This inherently implies that the OPR of the inherited ice is not wholly preserved. It is possible that due to beam smearing multiple components are hidden in what now seems to be a single component. However, the OPR drops across three beam sizes making this scenario unlikely.

If, on the other hand, we accept that the OPR is reset to 3.0 on desorption as suggested by the experiments discussed above, the lower OPR values found outside 60 au mean that ice formation cannot play a significant role here. Instead, gas formation is required. To explain the low OPR requires either low temperature formation or a \emph{chemical} explanation. The former can be explained by deeper penetration of UV radiation at large radii, producing the required radicals closer to the midplane and at lower temperatures. The latter requires detailed modeling including the spin state chemistry, and the role of H$_2$ spin.  The extent to which the \emph{rotational} temperature in the outer disk deviates from the \emph{spin} temperature corresponding to the observed OPR cannot be assessed with the current data. Future ALMA observations with higher signal-to-noise and additional transitions with lower upper state energies are needed for this.

Additionally, in this scenario the observed OPR of 3.0 inside 60~au is both consistent with gas-phase formation and non-thermal desorption from the solid-state. However, \ch{H2CO} formed in the solid-state from CO hydrogenation during the protoplanetary disk stage requires vertical transport which is unlikely in TW~Hya due to the lack of turbulence, as described above. This raises the question, what does create the \ch{H2CO} ring emission at $\sim$~20~au if it is not linked to the CO snowline? The observations of the edge-on younger embedded disk IRAS~04302 also find that \ch{H2CO} decreases in the inner region \citep{2020_Merel_emb_disks_arXiv200808106V, 2020_Podio_emb_disks_arXiv200812648P}. In this younger and warmer disk CO does not freeze-out due to higher midplane temperatures \citep{2018_Merel_L1527_A&A...615A..83V}. Furthermore, \ch{C^{17}O} emission in the IRAS~04302 disk does not decrease in the inner region, ruling out dust opacity \citep{2020_Merel_emb_disks_arXiv200808106V}. The authors thus argue that the decrease of \ch{H2CO} in the inner region is due to lower abundances of parent radicals in the gas-phase instead of an optically thick continuum. This mechanism could still be at play in an older protoplanetary disk like TW~Hya and will be investigated in a follow-up paper with chemical modelling.

\section{Summary} \label{sec:sum}
We report the most comprehensive survey of spatially and spectrally resolved ortho and para \ch{H2CO} emission in a protoplanetary disk to date, TW~Hya. We detect \ch{H2CO} emission across the entire disk out to 180~au, with a partially filled emission ring at 20~au and a smooth decrease beyond this radius. A rotational diagram analysis shows that the emission originates from a layer with a nearly constant temperature between 30 to 40~K, which corresponds to $z/R\ge$ 0.25. We find column densities of a few times $10^{13}$ cm$^{-2}$ in the inner disk decreasing to $\sim 10^{12}$ cm$^{-2}$ in the outer disk, and an OPR consistent with 3 in the inner 60~au decreasing to a value of $\sim 2$ at 120~au. Unlike some other disks, e.g., HD~163296, CI~Tau, DM~Tau, and AS~209, no secondary increase in the \ch{H2CO} emission or column density is seen in the outer disk. The results and discussion presented in this work lead us to speculate that the low OPR of \ch{H2CO} in the disk of TW Hya does not reflect direct ice-formation, as is commonly assumed, but instead hints at predominantly gas-phase formation. Several lines of evidence lead to this speculation: 1) the smooth emission profiles that suggests a single formation path across the disk, 2) the radially decreasing OPR, 3) the lack of vertical mixing to return \ch{H2CO} ice from the disk midplane, and 4) the recent results on the reset of the OPR to 3 upon desorption of \ch{H2O}. Instead, a cold \emph{gas-phase} origin of the gaseous \ch{H2CO} molecules responsible for the emission appears a more likely scenario or TW Hya. In other disks \citep[e.g., DM~Tau,][]{2015_Loomis_DMTAU_ApJ...809L..25L}, ice formation may play a larger role, and even in TW Hya the \emph{bulk} of the \ch{H2CO} likely resides (unobserved) in ice near the midplane. Gas-phase formation is supported by the presence of abundant \ch{H2CO} in the same region where there is a deficit of solid mass, specifically outside of the millimeter pebble disk. This is the same region where the OPR begins to drop. This scenario will be tested in a follow-up study with forward models that include chemistry, spin-states, and radiative transfer to better understand the observed OPR and its implications for organic formation in disks during planet formation.

\acknowledgments
The authors thank the anonymous referee for the constructive feedback on this manuscript. The authors acknowledge the help with the ALMA data processing by Allegro, the European ALMA Regional Center node in the Netherlands; Allegro is funded by NWO, the Netherlands Organisation for Scientific Research. This paper makes use of the following ALMA data:

\begin{itemize}
    \item ADS/JAO.ALMA\#2013.1.00114.S, 
    \item ADS/JAO.ALMA\#2016.1.00311.S,
    \item ADS/JAO.ALMA\#2016.1.00464.S.
\end{itemize}

ALMA is a partnership of ESO (representing its member states), NSF (USA) and NINS (Japan), together with NRC (Canada), MOST and ASIAA (Taiwan), and KASI (Republic of Korea), in cooperation with the Republic of Chile. The Joint ALMA Observatory is operated by ESO, AUI/NRAO and NAOJ. The National Radio Astronomy Observatory is a facility of the National Science Foundation operated under cooperative agreement by Associated Universities, Inc. J.T.v.S. and M.R.H. are supported by the Dutch Astrochemistry II program of the Netherlands Organization for Scientific Research (648.000.025). L.I.C. gratefully acknowledges support from the David and Lucille Packard Foundation, the VSGC New Investigators Award, and NASA ATP 80NSSC20K0529. C.W.~acknowledges financial support from the University of Leeds and from the Science and Technology Facilities Council (grant numbers ST/R000549/1 and ST/T000287/1). J.K.C. acknowledges support from the National Science Foundation Graduate Research Fellowship under Grant No. DGE 1256260 and the National Aeronautics and Space Administration FINESST grant, under Grant no. 80NSSC19K1534. V.V.G. acknowledges support from FONDECYT Iniciación 11180904. J.H. acknowledges support for this work provided by NASA through the NASA Hubble Fellowship grant \#HST-HF2-51460.001-A awarded by the Space Telescope Science Institute, which is operated by the Association of Universities for Research in Astronomy, Inc., for NASA, under contract NAS5-26555."M.K. gratefully acknowledges funding by the University of Tartu ASTRA project 2014-2020.4.01.16-0029 KOMEET, financed by the EU European Regional Development Fund.

%








\bibliographystyle{aasjournal}



\end{document}